\documentclass[useAMS,usegraphicx]{mn2e}

\newcommand{\Msun}{\ensuremath{\,{\rm M}_\odot}}                        
\newcommand{\Rsun}{\ensuremath{\,{\rm R}_\odot}}                        
\newcommand{\Teff}{\ensuremath{T_{\rm eff}}}                            
\newcommand{\logg}{\ensuremath{\log g}}                                 
\newcommand{\Veq}{\ensuremath{V_{\rm eq}}}                              
\newcommand{\kms}{\,km\,s$^{-1}$}                                       
\newcommand{\ion}[2]{{#1}\,{\sc {\small{#2}}}}                          
\newcommand{\cms}{\,cm\,s$^{-1}$}                                       
\newcommand{\Vsync}{\ensuremath{V_{\rm synch}}}                         
\newcommand{\FeH}{\ensuremath{\left[\frac{\rm Fe}{\rm H}\right]}}       
\newcommand{\MoH}{\ensuremath{\left[\frac{\rm M}{\rm H}\right]}}        

\title[Absolute dimensions of eclipsing binary WW\,Aur]
      {Absolute dimensions of detached eclipsing binaries. \\ I. The metallic-lined system WW\,Aurigae}

\author[J.\ Southworth et al.]
       {J.\ Southworth$^1$\thanks{E-mail: jkt@astro.keele.ac.uk (JS), pflm@astro.keele.ac.uk \newline (PFLM), 
        bs@astro.keele.ac.uk (BS), claret@iaa.es (AC) and \newline etzel@mintaka.sdsu.edu (PBE)}, B.\ Smalley$^1$\footnotemark[1] 
        P.\ F.\ L.\ Maxted$^1$\footnotemark[1], A.\ Claret$^2$\footnotemark[1] and P.\ B.\ Etzel$^3$\footnotemark[1] \\
        $^1$\,Department of Physics and Chemistry, Keele University, Staffordshire, ST5 5BG, UK \\
        $^2$\,Instituto de Astrof\'\i sica de Andaluc\'\i a, CSIC, Apartado 3004, 18080 Granada, Spain \\
        $^3$\,Department of Astronomy, San Diego State University, San Diego, CA 92182, USA}

\begin{document} \maketitle 

\begin{abstract}
WW\,Aurigae is a detached eclipsing binary composed of two metallic-lined A-type stars orbiting each other every 2.5 days. We have determined the masses and radii of both components to accuracies of 0.4\% and 0.6\%, respectively. From a cross-correlation analysis of high-resolution spectra we find masses of $1.964 \pm 0.007$\Msun\ for the primary star and $1.814 \pm 0.007$\Msun\ for the secondary star. From an analysis of photoelectric $uvby$ and $UBV$ light curves we find the radii of the stars to be $1.927 \pm 0.011$\Rsun\ and $1.841 \pm 0.011$\Rsun, where the uncertainties have been calculated using a Monte Carlo algorithm. Fundamental effective temperatures of the two stars have been derived, using the Hipparcos parallax of WW\,Aur and published ultraviolet, optical and infrared fluxes, and are $7960 \pm 420$ and $7670 \pm 410$\,K. The masses, radii and effective temperatures of WW\,Aur are only matched by theoretical evolutionary models for a fractional initial metal abundance, $Z$, of approximately 0.06 and an age of roughly 90\,Myr. This seems to be the highest metal abundance inferred for a well-studied detached eclipsing binary, but we find no evidence that it is related to the metallic-lined nature of the stars. The circular orbit of WW\,Aur is in conflict with the circularization timescales of both the Tassoul and the Zahn tidal theories and we suggest that this is due to pre-main-sequence evolution or the presence of a circular orbit when the stars were formed.
\end{abstract}



\section{Introduction}                                                                                \label{sec:intro}    

The analysis of photometric and spectroscopic observations of detached eclipsing binary stars (dEBs) is one of the most powerful ways to investigate the properties of single stars (Andersen 1991). The analysis of good light curves and radial velocity curves allows the measurement of the absolute masses and radii of two stars to an accuracy of 1\% or better, and surface gravities to within 0.01\,dex (e.g., Southworth, Maxted \& Smalley 2004b). Further analysis using spectral synthesis or photometric index calibrations also allows the derivation of accurate effective temperatures and therefore luminosities of the components of the eclipsing system. 

Accurate dimensions of a dEB provide a way of testing theoretical stellar evolutionary models because the two components of the system have the same age and chemical composition but, in general, different masses and radii. The observation and analysis of dEBs in stellar clusters and associations may also allow the calculation of accurate dimensions of four or more stars of the same age and initial chemical composition. In this case the distance to the cluster can be found using empirical methods which are free of the difficulties of the main-sequence fitting technique (Southworth, Maxted \& Smalley 2005). Furthermore, the age, metallicity and helium abundance of the cluster as a whole can be found from comparison of the observed masses and radii of the dEB with theoretical stellar models (Southworth, Maxted \& Smalley 2004a).

Another use of dEBs is to investigate the physical processes at work in single stars. This is particularly useful if one or both components belongs to a class of peculiar or poorly understood stars, for example slowly pulsating B\,stars (e.g., V539\,Arae; Clausen 1996), or metallic-lined stars (e.g., KW\,Hydrae; Andersen \& Vaz 1984, 1987). Metallic-lined A stars (Am stars) are well represented in the compilation of accurate eclipsing binary data by Andersen (1991), but many details of why and how these stars form and evolve are still not well understood.

Am stars have spectral types between late-B and early-F, and exhibit peculiar photospheric chemical abundances. Whilst the calcium and scandium spectral lines are abnormally weak, the lines due to other metals can be strongly enhanced. This phenomenon is observed to be common for A stars with rotational velocities below about 100\kms\ (Abt \& Morrell 1995), and is caused by the diffusion of chemical species below the surfaces of the stars (Budaj 1996, 1997; Michaud et al.\ 2004; Turcotte et al.\ 2000). Am stars often occur in binary systems (e.g., Abt 1961), which is expected as tidal effects in close binaries generally cause the rotational velocities of the stars to be slower than those of single stars. Whilst there have been suggestions that Am stars have slightly larger radii than other A-type stars (Kitamura \& Kondo 1978; Budaj 1996), this has not been confirmed by recent studies of metallic-lined dEBs (Lacy et al.\ 2002, 2004a; Lacy, Claret \& Sabby 2004b). The masses, radii and luminosities of metallic-lined components of dEBs are generally well matched by the predictions of theoretical evolutionary models for normal stars.

\begin{table} \begin{center} 
\caption{\label{table:wwaurdata} Identifications, location, and combined photometric indices for WW\,Aurigae.
\newline {\bf References:} (1) Perryman et al.\ (1997); (2) Cannon \& Pickering (1918); (3) Peters \& Hoffleit (1992); (4) Argelander (1903); (5) KK75; (6) 2MASS; (7) Crawford et al.\ (1972).}                        
\begin{tabular}{lr@{}lr} \hline
                        &     & WW Aurigae            & Reference \\ \hline
Hipparcos number        &     & HIP 31173             & 1   \\
Henry Draper number     &     & HD 46052              & 2   \\
Bright Star Catalogue   &     & HR 2372               & 3   \\
Bonner Durchmusterung   &     & BD\,+32\degr 1324     & 4   \\[3pt]
$\alpha_{2000}$         &     & 06 32 27.19           & 1   \\
$\delta_{2000}$         &   + & 32 27 17.6            & 1   \\
Hipparcos parallax ($m$as) &  & $11.86 \pm 1.06$      & 1   \\
Spectral type           &     & A4\,m + A5\,m         & 5   \\[3pt]
$B_T$                   &     & 6.036 $\pm$ 0.005     & 1   \\
$V_T$                   &     & 5.839 $\pm$ 0.005     & 1   \\
$J_{\rm 2MASS}$         &     & 5.498 $\pm$ 0.021     & 6   \\
$H_{\rm 2MASS}$         &     & 5.499 $\pm$ 0.026     & 6   \\
$K_{\rm 2MASS}$         &     & 5.481 $\pm$ 0.021     & 6   \\[3pt]
$b-y$                   &   + & 0.081 $\pm$ 0.008     & 7   \\
$m_1$                   &   + & 0.231 $\pm$ 0.011     & 7   \\
$c_1$                   &   + & 0.944 $\pm$ 0.011     & 7   \\
$\beta$                 &     & 2.862 $\pm$ 0.013     & 7   \\
\hline \end{tabular} \end{center} \end{table}

\subsection{WW Aurigae}

WW\,Aurigae ($P = 2.52$ days, $m_V = 5.8$) is a bright Northern hemisphere metallic-lined dEB (Table~\ref{table:wwaurdata}). It has an accurate trigonometrical parallax from the Hipparcos satellite, which gives a distance of $84.3 \pm 7.6$\,pc. Its eclipsing nature was discovered independently by Soloviev (1918) and Schwab (1918). Joy (1918) presented spectra which showed lines of both components moving with an orbital period of 2.525 days. Wylie (1930) verified this period photometrically. Dugan (1930) made extensive photometric observations but the investigation was complicated by the slight photometric variability of the comparison star. Huffer \& Kopal (1951) and Piotrowski \& Serkowski (1956) undertook photoelectric observations of WW\,Aur but were both hampered by bright observing conditions. 

Etzel (1975, hereafter E75) observed excellent photoelectric light curves in the Str\"omgren $uvby$ passbands, consisting of about one thousand observations in each passband. The light curve analysis code {\sc ebop} (see Section~\ref{sec:ebop}) was introduced in E75 and used to derive the photometric elements of WW\,Aur from the $uvby$ light curves.

Kiyokawa \& Kitamura (1975, hereafter KK75) published excellent $UBV$ light curves of WW\,Aur and analysed them using a procedure based on rectification (Russell \& Merrill 1952; Kitamura 1967). Kitamura, Kim \& Kiyokawa (1976) published good photographic spectra and derived accurate absolute dimensions of WW\,Aur by combining their results with those of KK75. The KK75 light curves were also analysed by Cester et al.\ (1978) in their program to determine accurate and homogeneous photometric elements of many dEBs with the light curve modelling program {\sc wink} (Wood 1971).

We shall refer to the hotter and more massive star as component A and the cooler and less massive star as component B. Abt \& Morrell (1995) classified both stars as Am (A2,A5,A7) where the bracketed spectral types have been obtained from the \ion{Ca}{II}\,K line, Balmer lines, and metallic lines, respectively.


\section{Observations and data aquisition}                                                            \label{sec:obs}

\subsection{Spectroscopy}

Spectroscopic observations were carried out in 2002 October using the 2.5\,m Isaac Newton Telescope (INT) on La Palma. The 500\,mm camera of the Intermediate Dispersion Spectrograph (IDS) was used with a holographic 2400\,{\it l}\,mm$^{-1}$ grating. An EEV 4k\,$\times$\,2k CCD was used and exposure times were 120 seconds. From measurements of the full width at half maximum (FWHM) of arc lines taken for wavelength calibration we estimate that the resolution is approximately 0.2\,\AA\ (corresponding to two pixels). The spectral window chosen for observation was 4230--4500\,\AA. This contains the \ion{Mg}{II}\ 4481\,\AA\ line which is known to be one of the best lines for radial velocity work for early-type stars (Andersen 1975). A total of 59 spectra were obtained with an estimated signal to noise ratio per pixel of about 450. One spectrum, with a signal to noise ratio of about 700, was observed at H$\beta$ (4861\,\AA) to provide an additional indicator of the atmospheric properties of the stars. 

A wide range of standard stars were observed during the same observing run and using the same observing equipment. The reduction of all spectra was undertaken using optimal extraction as implemented in the software tools {\sc pamela} and {\sc molly} \footnote{{\sc pamela} and {\sc molly} were written by Dr.\ Tom Marsh and are available at {\tt http://www.warwick.ac.uk/staff/T.R.Marsh}} (Marsh 1989).

\subsection{Photometry}

Photoelectric $uvby$ light curves were obtained by PBE between 1973 September and 1974 April using a 41\,cm Cassegrain telescope at Mt.\ Laguna Observatory, USA, a single-channel photometer with a refrigerated 1P21 photomultiplier, and a 36\,arcsecond diaphragm (E75). Observations were taken in the sequence variable--comparison--sky with integration times of 15 seconds. The comparison star (HD\,46251, spectral type A2\,V) was compared to a check star (HD\,48272) and no variability in brightness was found. The $uvby$ light curves are contained in Tables \ref{table:ulc}, \ref{table:vlc}, \ref{table:blc} and \ref{table:ylc}, which will be given in full in the electronic edition of this work and in an archive at the CDS\footnote{\tt http://cdsweb.u-strasbg.fr/}. They can also be obtained as file 44 from the IAU Archives of Unpublished Observations of Variable Stars (Breger 1982).

The photoelectric light curves observed by KK75 in the Johnson $UBV$ passbands each contain approximately 1000 observations. These are clearly tabulated and were converted to digital form using proprietary optical character recognition software. This can cause occasional errors (for example a 3 is easily mistaken for an 8), so the data have been inspected and all clear mistakes repaired. Some mistakes which are smaller than the observational errors may remain; we estimate that the 3000 observations from KK75 now contain fewer than ten misreadings. Interested researchers can obtain this data by contacting JS.

\begin{table} \begin{center} \caption{\label{table:ulc} Sample of the $u$ light curve of E75. The full table is given in the electronic edition of this work.}
\begin{tabular}{lr} \hline
HJD - 2\,400\,000 & Differential $u$ magnitude \\ \hline
41945.87344 & -0.52164 \\
41945.88766 & -0.47606 \\
41945.89511 & -0.42789 \\
41945.90211 & -0.42268 \\
41945.90783 & -0.39867 \\
41945.92127 & -0.29304 \\
41945.93044 & -0.25070 \\
41945.93686 & -0.18481 \\
41945.94392 & -0.16004 \\
41945.94999 & -0.12016 \\ \hline
\end{tabular} \end{center} \end{table}

\begin{table} \begin{center} \caption{\label{table:vlc} Sample of the $v$ light curve of E75. The full table is given in the electronic edition of this work.}
\begin{tabular}{lr} \hline
HJD - 2\,400\,000 & Differential $v$ magnitude \\ \hline
41945.87291 & -0.60037 \\
41945.88033 & -0.56323 \\
41945.88713 & -0.56664 \\
41945.89443 & -0.53120 \\
41945.90151 & -0.50243 \\
41945.90729 & -0.46397 \\
41945.91339 & -0.42306 \\
41945.92077 & -0.37732 \\
41945.92993 & -0.31512 \\
41945.93619 & -0.26960 \\ \hline
\end{tabular} \end{center} \end{table}

\begin{table} \begin{center} \caption{\label{table:blc} Sample of the $b$ light curve of E75. The full table is given in the electronic edition of this work.}
\begin{tabular}{lr} \hline
HJD - 2\,400\,000 & Differential $b$ magnitude \\ \hline
41945.86559 & -0.47952 \\
41945.87316 & -0.48512 \\
41945.88738 & -0.46660 \\
41945.89473 & -0.39264 \\
41945.90182 & -0.39278 \\
41945.90754 & -0.35846 \\
41945.91363 & -0.28938 \\
41945.92103 & -0.24824 \\
41945.93015 & -0.21623 \\
41945.93642 & -0.16705 \\ \hline
\end{tabular} \end{center} \end{table}

\begin{table} \begin{center} \caption{\label{table:ylc} Sample of the $y$ light curve of E75. The full table is given in the electronic edition of this work.}
\begin{tabular}{lr} \hline
HJD - 2\,400\,000 & Differential $y$ magnitude \\ \hline
41945.87236 & -0.63222 \\
41945.88009 & -0.61147 \\
41945.88689 & -0.58301 \\
41945.89399 & -0.56776 \\
41945.90122 & -0.53937 \\
41945.90701 & -0.51724 \\
41945.91317 & -0.45826 \\
41945.92045 & -0.39762 \\
41945.92971 & -0.35776 \\
41945.93592 & -0.31868 \\ \hline
\end{tabular} \end{center} \end{table}


\section{Period determination}                                                                        \label{sec:period}     

\begin{table} \begin{center} 
\caption{\label{table:minima} Literature times of minimum light of WW\,Aurigae and the observed minus calculated ($O-C$) values of the data compared to the ephemeris derived in this work.
\newline $^\dag$\,Rejected from the solution due to large $O-C$ value.
\newline {\bf References:} (1) Huffer \& Kopal (1951); (2) Piotrowski \& Serkowski (1956); (3) Fitch (1964); (4) Broglia \& Lenouvel (1960); (5) Chou (1968); (6) D.\ B.\ Wood (1973, private communication); (7) Kristenson (1966); (8) Pohl \& Kizilirmak (1966); (9) Robinson \& Ashbrook (1968); (10) KK75; (11) Baldwin (1973); (12) Popovici (1968); (13) Pohl \& Kizilirmak (1970); (14) H.\ Lanning (E75); (15) Popovici (1971); (16) Pohl \& Kizilirmak (1972); (17) Kizilirmak \& Pohl (1974); (18) E75; (19) Ebersberger, Pohl \& Kizilirmak (1978); (20) Pohl et al.\ (1982); (21) Caton, Burns \& Hawkins (1991).} 
\begin{tabular}{lrr@{}lr} \hline
Time of minimum         & Cycle     & \multicolumn{2}{r}{$O-C$ value}   & Reference       \\ 
(HJD $-$ 2\,400\,000)   & number    & \hspace{20pt} (HJD)   &           &                 \\ \hline
32480.9379              &$-$3758.0  &   0.0025  &         &  1  \\
32868.5250              &$-$3604.5  &$-$0.0009  &         &  2  \\
32888.7263              &$-$3596.5  &   0.0002  &         &  1  \\
32892.5129              &$-$3595.0  &$-$0.0007  &         &  2  \\
32936.6998              &$-$3577.5  &$-$0.0016  &         &  1  \\
32945.5389              &$-$3574.0  &$-$0.0001  &         &  2  \\
32945.5403              &$-$3574.0  &   0.0013  &         &  2  \\
32946.8002              &$-$3573.5  &$-$0.0013  &         &  1  \\
33002.3510              &$-$3551.5  &$-$0.0009  &         &  2  \\
33031.3878              &$-$3540.0  &$-$0.0019  &         &  2  \\
33190.4670              &$-$3477.0  &   0.0011  &         &  2  \\
33209.4042              &$-$3469.5  &   0.0007  &         &  2  \\
33215.7165              &$-$3467.0  &   0.0004  &         &  1  \\
33225.8159              &$-$3463.0  &$-$0.0003  &         &  1  \\
33249.8035              &$-$3453.5  &$-$0.0003  &         &  1  \\
33263.6905              &$-$3448.0  &$-$0.0010  &         &  1  \\
33292.7299              &$-$3436.5  &   0.0007  &         &  1  \\
33297.7776              &$-$3434.5  &$-$0.0016  &         &  1  \\
33358.3816              &$-$3410.5  &   0.0019  &         &  2  \\
33570.4817              &$-$3326.5  &   0.0004  &         &  2  \\
33594.4695              &$-$3317.0  &   0.0005  &         &  2  \\
33599.5173              &$-$3315.0  &$-$0.0017  &         &  2  \\
33646.2338              &$-$3296.5  &   0.0019  &         &  2  \\
33690.4196              &$-$3279.0  &$-$0.0001  &         &  2  \\
34470.650               &$-$2970.0  &$-$0.0007  &         &  3  \\
35845.5222              &$-$2425.5  &$-$0.0016  &         &  4  \\
36586.616               &$-$2132.0  &$-$0.0010  &         &  5  \\
36591.6714              &$-$2130.0  &   0.0044  &         &  5  \\
\hline \end{tabular} \end{center} \end{table}
\begin{table}\begin{center}\contcaption{}
\begin{tabular}{lrr@{}lr} \hline
Time of minimum         & Cycle     & \multicolumn{2}{r}{$O-C$ value}   & Reference       \\ 
(HJD $-$ 2\,400\,000)   & number    & \hspace{20pt} (HJD)   &           &                 \\ \hline
37654.7002              &$-$1709.0  &$-$0.0000  &         &  6  \\
38793.4841              &$-$1258.0  &   0.0001  &         &  7  \\
38798.5335              &$-$1256.0  &$-$0.0005  &         &  7  \\
38802.3215              &$-$1254.5  &$-$0.0000  &         &  7  \\
38807.3713              &$-$1252.5  &$-$0.0003  &         &  7  \\
38831.3589              &$-$1243.0  &$-$0.0003  &         &  7  \\
39134.3610              &$-$1123.0  &$-$0.0006  &         &  8  \\
39184.8586              &$-$1103.0  &$-$0.0034  &         &  9  \\
39527.0020              & $-$967.5  &$-$0.0001  &         &  10  \\
39537.1022              & $-$963.5  &   0.0000  &         &  10  \\
39550.9895              & $-$958.0  &$-$0.0003  &         &  10  \\
39556.0398              & $-$956.0  &$-$0.0000  &         &  10  \\
39835.0550              & $-$845.5  &   0.0005  &         &  10  \\
39836.3167              & $-$845.0  &$-$0.0003  &         &  10  \\
39852.7308              & $-$838.5  &   0.0012  &         &  11  \\
39857.7790              & $-$836.5  &$-$0.0006  &         &  11  \\
39864.0920              & $-$834.0  &$-$0.0002  &         &  10  \\
39869.1424              & $-$832.0  &   0.0002  &         &  11  \\
39857.7790              & $-$836.5  &$-$0.0006  &         &  10  \\
39888.0810              & $-$824.5  &   0.0011  &         &  10  \\
40154.4692              & $-$719.0  &$-$0.0002  &         &  12  \\
40288.298               & $-$666.0  &   0.0026  &         &  13  \\
40684.7235              & $-$509.0  &   0.0000  &         &  14  \\
40885.4635              & $-$429.5  &   0.0010  &         &  15  \\
41024.3382              & $-$374.5  &$-$0.0004  &         &  16  \\
41399.305               & $-$226.0  &   0.0010  &         &  17  \\
41945.9707              &   $-$9.5  &   0.0000  &         &  18  \\
41969.9585              &      0.0  &   0.0001  &         &  18  \\
41983.8458              &      5.5  &$-$0.0002  &         &  18  \\
42022.9841              &     21.0  &   0.0003  &         &  18  \\
42026.7718              &     22.5  &   0.0005  &         &  18  \\
42028.0342              &     23.0  &   0.0004  &         &  18  \\
42069.6970              &     39.5  &   0.0004  &         &  18  \\
42103.7847              &     53.0  &   0.0003  &         &  18  \\
42117.6725              &     58.5  &   0.0005  &         &  18  \\
42141.6602              &     68.0  &   0.0005  &         &  18  \\
43477.3949              &    597.0  &$-$0.0001  &         &  19  \\
44256.3632              &    905.5  &$-$0.0002  &         &  20  \\
44925.4902              &   1170.5  &$-$0.0034  &         &  20  \\
46840.73088$^\dag$      &   1929.0  &   0.0112  & $^\dag$ &  21  \\
\hline \end{tabular} \end{center} \end{table}

\begin{figure*} \includegraphics[width=\textwidth,angle=0]{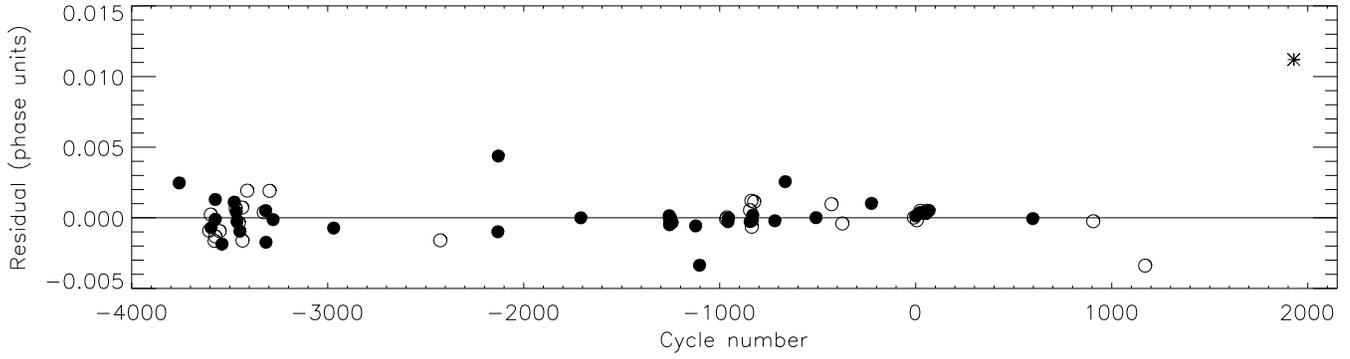} \\ \caption{\label{fig:periodresids} 
Residuals (in units of the orbital period) of the ephemeris which best fits the observed times of minima. The filled circles refer to primary minima and the open circles to secondary minima. The starred symbol at cycle 1929.0 represents a datapoint which was rejected from the fit.} \end{figure*}

Photoelectric times of minima were collected from the literature, and the orbital ephemeris given in the {\it General Catalogue of Variable Stars} (Khopolov et al.\ 1999) was used to determine the preliminary cycle number of each minimum (reference time of minimum HJD 2\,432\,945.5393 and period 2.52501922 days). Equal weights were given to all observations as very few have quoted uncertainties. A straight line was fitted to the resulting cycle numbers and times of minima (Table~\ref{table:minima}) by $\chi^2$ minimisation, using the first time of primary minimum from E75 as the reference time. One time of minimum had a large residual so was rejected. 

The resulting ephemeris is: \[ {\rm Min\,I} = {\rm HJD}\ 2\,441\,969.95837(23) + 2.52501941(10) \times E \]
where each number in parentheses denotes the uncertainty in the final digit of the preceding number. All uncertainties quoted in this work are standard errors unless otherwise stated. The residuals of the fit are plotted in Fig.~\ref{fig:periodresids} and give no indication of any form of period change. The root-mean-square of the residuals is 0.0012 days. The above ephemeris is in good agreement with that given in the {\it Eclipsing Binaries Minima Database}\footnote{\tt http:://www.as.ap.krakow.pl/} (Kreiner, Kim \& Nha 2001).


\section{Spectroscopic orbits}                                                                              \label{sec:todcor}   

\begin{table} \begin{center} \caption{\label{table:rvs} Radial velocities for WW\,Aur calculated using {\sc todcor} with template spectra from HD\,39945 (for star A) and HD\,32115 (for star B). These velocities are only a small part of the information used to calculate the final spectroscopic orbits for the components of WW\,Aur (see text for details). The velocities in this table are relative to the observed template spectra; they can be brought into the heliocentric rest frame by adding 8.6 and 38.2\kms\ for star A and star B, respectively.}
\begin{tabular}{lrrrrr} \hline
HJD $-$     & Primary  & $O-C$ & Secondary  & $O-C$ & Wt \\ 
2\,400\,000 & velocity &       & velocity   &       &    \\ \hline
52568.7191 &  -22.7 &  -3.3 &  -44.2 &   0.5 & 0.7 \\
52569.5637 &   82.7 &  -2.4 & -158.5 &  -0.8 & 1.3 \\
52569.5653 &   83.3 &  -1.6 & -157.6 &  -0.1 & 1.6 \\
52569.5669 &   83.2 &  -1.5 & -156.8 &   0.5 & 1.8 \\
52569.6347 &   70.8 &  -2.7 & -146.4 &  -1.2 & 1.1 \\
52569.6982 &   58.1 &  -2.6 & -131.6 &  -0.3 & 1.6 \\
52569.7304 &   51.4 &  -2.0 & -124.7 &  -1.2 & 1.4 \\
52569.7320 &   51.4 &  -1.7 & -123.8 &  -0.7 & 1.2 \\
52570.5994 & -136.8 &  -2.8 &   77.1 &  -2.2 & 1.2 \\
52570.6010 & -132.5 &   1.5 &   82.1 &   2.8 & 0.7 \\
52570.6026 & -134.2 &  -0.1 &   80.1 &   0.7 & 1.1 \\
52570.6041 & -133.9 &   0.1 &   80.4 &   1.0 & 1.1 \\
52570.6057 & -134.5 &  -0.4 &   79.5 &   0.1 & 1.1 \\
52570.6550 & -134.8 &  -1.0 &   78.9 &  -0.2 & 1.4 \\
52570.6566 & -133.9 &  -0.3 &   80.1 &   1.1 & 1.2 \\
52570.6582 & -133.4 &   0.3 &   80.4 &   1.4 & 1.2 \\
52570.7404 & -129.3 &  -0.2 &   74.9 &   0.9 & 1.0 \\
52570.7420 & -129.9 &  -1.0 &   74.3 &   0.5 & 1.3 \\
52570.7435 & -129.6 &  -0.8 &   74.0 &   0.3 & 1.5 \\
52570.7781 & -124.4 &   1.0 &   71.4 &   1.4 & 1.6 \\
52570.7797 & -123.5 &   1.6 &   71.4 &   1.6 & 1.4 \\
52570.7813 & -122.9 &   2.1 &   71.4 &   1.8 & 1.2 \\
52570.7878 & -120.9 &   3.4 &   71.7 &   2.9 & 0.5 \\
52570.7894 & -120.9 &   3.2 &   72.2 &   3.7 & 0.4 \\
52570.7910 & -123.8 &   0.1 &   67.9 &  -0.4 & 1.2 \\
52571.5705 &   63.5 &  -2.8 & -137.4 &  -0.0 & 0.8 \\
\hline \end{tabular} \end{center} \end{table}
\begin{table}\begin{center}\contcaption{}
\begin{tabular}{lrrrrr} \hline
HJD $-$     & Primary  & $O-C$ & Secondary  & $O-C$ & Wt \\ 
2\,400\,000 & velocity &       & velocity   &       &    \\ \hline
52571.5721 &   65.0 &  -1.6 & -136.8 &   0.9 & 0.9 \\
52571.5737 &   66.5 &  -0.5 & -136.2 &   1.8 & 0.6 \\
52571.6700 &   86.1 &   2.2 & -152.4 &   4.0 & 0.8 \\
52571.6716 &   85.8 &   1.7 & -153.6 &   3.1 & 0.6 \\
52571.6732 &   87.0 &   2.6 & -152.2 &   4.8 & 0.4 \\
52571.7428 &   95.1 &   2.2 & -165.2 &   1.0 & 1.1 \\
52571.7444 &   94.8 &   1.7 & -165.8 &   0.6 & 1.0 \\
52571.7460 &   95.1 &   1.9 & -166.1 &   0.5 & 1.2 \\
52571.7732 &   96.0 &   0.3 & -170.1 &  -0.9 & 1.4 \\
52571.7748 &   96.3 &   0.5 & -170.2 &  -0.9 & 1.2 \\
52571.7764 &   96.9 &   1.0 & -169.8 &  -0.4 & 1.3 \\
52571.7779 &   97.5 &   1.4 & -170.1 &  -0.5 & 1.2 \\
52571.7795 &   97.8 &   1.6 & -170.2 &  -0.5 & 1.1 \\
52571.7813 &   98.0 &   1.7 & -169.6 &   0.3 & 0.8 \\
52571.7829 &   98.6 &   2.1 & -169.3 &   0.7 & 0.8 \\
52571.7845 &   98.9 &   2.3 & -169.6 &   0.5 & 0.6 \\
52571.7861 &   98.9 &   2.2 & -169.6 &   0.6 & 0.5 \\
52571.7877 &   98.9 &   2.1 & -169.8 &   0.5 & 0.4 \\
52572.6408 &  -52.6 &   0.6 &  -10.6 &  -2.4 & 0.9 \\
52572.6424 &  -52.9 &   0.7 &  -10.0 &  -2.3 & 0.7 \\
52572.6439 &  -53.7 &   0.3 &   -9.4 &  -2.2 & 0.9 \\
52572.7501 &  -81.5 &   0.2 &   22.4 &  -0.3 & 0.4 \\
52572.7517 &  -81.8 &   0.3 &   22.7 &  -0.5 & 0.4 \\
52572.7533 &  -82.1 &   0.4 &   23.3 &  -0.3 & 0.3 \\
52572.7846 &  -89.3 &   0.5 &   28.8 &  -2.8 & 0.3 \\
52572.7862 &  -90.2 &   0.0 &   28.8 &  -3.2 & 0.6 \\
52572.7878 &  -90.2 &   0.4 &   29.4 &  -3.0 & 0.5 \\
52572.7895 &  -91.0 &  -0.1 &   28.8 &  -4.0 & 0.4 \\
52572.7911 &  -90.7 &   0.6 &   29.7 &  -3.5 & 0.6 \\
52572.7928 &  -90.7 &   1.0 &   30.3 &  -3.3 & 0.3 \\
52572.7944 &  -92.2 &  -0.1 &   29.7 &  -4.3 & 0.6 \\
52572.7960 &  -92.5 &  -0.1 &   30.0 &  -4.4 & 0.4 \\
\hline \end{tabular} \end{center} \end{table}

\begin{table*} \caption{\label{table:rvall} Radial velocities for the two components of WW\,Aur calculated using {\sc todcor} with different template stars. Radial velocities for the fifteen good template combinations are given. The radial velocities are relative to the templates and can be converted to the heliocentric rest frame by adding $-12.5$\kms\ (HD\,172167), 38.2\kms\ (HD\,32115), 19.9\kms\ (HD\,37594), 8.6\kms\ (HD\,39945), $-8.0$\kms\ (HD\,4222), $-29.6$\kms\ (HD\,905) or 31.2\kms\ (HR\,1218).}
{\em This table will be available in the electronic edition of the paper}
\end{table*}

Radial velocities were measured from the observed spectra using the two-dimensional cross-correlation algorithm {\sc todcor} (Zucker \& Mazeh 1994). In this algorithm two template spectra are simultaneously cross-correlated against each observed spectrum for a range of possible velocities for each template. The metallic-lined nature of both stars means that care must be taken to select template spectra which match the true spectra of the stars as closely as possible. For this reason synthetic spectra may not be the best choice as undetectable systematic errors could occur from missing or poorly matching spectral lines. 

Seven spectra, with spectral types between A0 and F5 (luminosity classes V and IV), were selected as templates from our set of standard star observations. The H$\gamma$ 4340\,\AA\ line was masked in all spectra during the radial velocity analysis. Radial velocities were calculated using {\sc todcor} for every combination of two template spectra, and are given in Table~\ref{table:rvall} in the electronic version of this work. Single-lined spectroscopic orbits were fitted to each set of radial velocities for the two stars using the {\sc sbop}\footnote{Spectroscopic Binary Orbit Program, \\ {\tt http://mintaka.sdsu.edu/faculty/etzel/}} program (Etzel 2004), which is a modified and expanded version of an earlier code by Wolfe, Horak \& Storer (1967). Initial investigations revealed that the orbital eccentricity is negligible so circular orbits were assumed. 

\begin{table} \begin{center} \caption{\label{table:rvorbit} Parameters of the spectroscopic orbit derived for WW\,Aur using {\sc todcor} with observed standard star spectra for templates. The systemic velocities were calculated using synthetic template spectra.}
\begin{tabular}{l r@{\,$\pm$\,}l r@{\,$\pm$\,}l} \hline
                                          & \multicolumn{2}{c}{Primary} & \multicolumn{2}{c}{Secondary}     \\ \hline
Semiamplitude $K$ (\kms)                  &          116.81 &  0.23     &          126.49 &  0.28           \\
Systemic velocity (\kms)                  &         $-$9.10 &  0.25     &         $-$7.84 &  0.32           \\
Mass ratio $q$                            &           \multicolumn{4}{c}{0.9235 $\pm$ 0.0027}               \\
$a \sin i$ (\Rsun)                        &           \multicolumn{4}{c}{12.138 $\pm$ 0.018}                \\
$M \sin^3 i$ (\Msun)                      &           1.959 & 0.007     &           1.809 & 0.007           \\
\hline \end{tabular}\end{center} \end{table}

\begin{figure} \includegraphics[width=0.48\textwidth,angle=0]{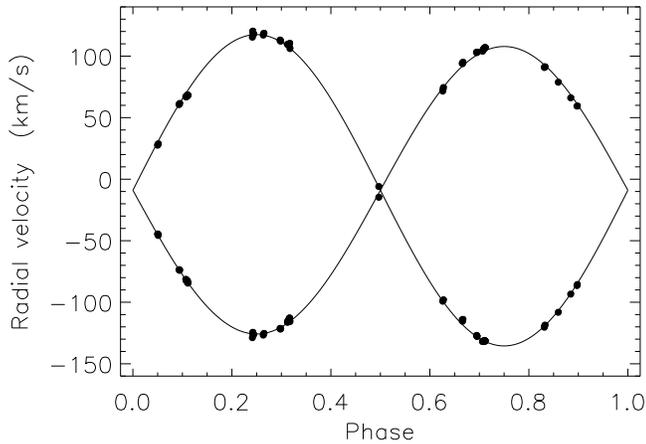} \\ \caption{\label{fig:rvorbit} 
Spectroscopic orbit for WW\,Aur from an {\sc sbop} fit to radial velocities from {\sc todcor}. The radial velocities are from Table~\ref{table:rvs} but have been adjusted to match the systemic velocities found using synthetic template spectra (see text for details).} \end{figure}

Good spectroscopic orbits were found for fifteen combinations of two template spectra; the remaining orbits are in good agreement but with larger standard errors. The final spectroscopic orbital parameters are given in Table~\ref{table:rvorbit} and are the mean and standard deviation of the parameters of the fifteen good individual orbits for the two stars. This procedure should have minimised possible systematic errors due to differences in effective temperature, metal abundance and rotational velocity between the template stars and the components of WW\,Aur. The standard deviation of the parameters of the final orbits are in excellent agreement with the estimated standard errors calculated by {\sc sbop} for each individual orbit.

An example spectroscopic solution has been selected from the solutions for the fifteen good combinations of template spectra. This solution was obtained using HD\,39945 (spectral type A5\,V) and HD\,32115 (A8\,IV) as templates for star A and star B, respectively. The radial velocities for this solution have been given in Table~\ref{table:rvs} and compared in Fig.~\ref{fig:rvorbit} with the final spectroscopic orbits for the components of WW\,Aur.

The systemic velocity of WW\,Aur have been found by calculating spectroscopic orbits using {\sc todcor} and synthetic template spectra constructed using {\sc uclsyn} (see Section~\ref{sec:teffs}). Separate systemic velocities were measured for each star (see Popper \& Hill 1991 for reasons); the poor agreement between the values for the two stars is probably caused by their spectral peculiarity.


\section{Light curve analysis}                                                                        \label{sec:ebop}

\begin{figure*} \includegraphics[width=\textwidth,angle=0]{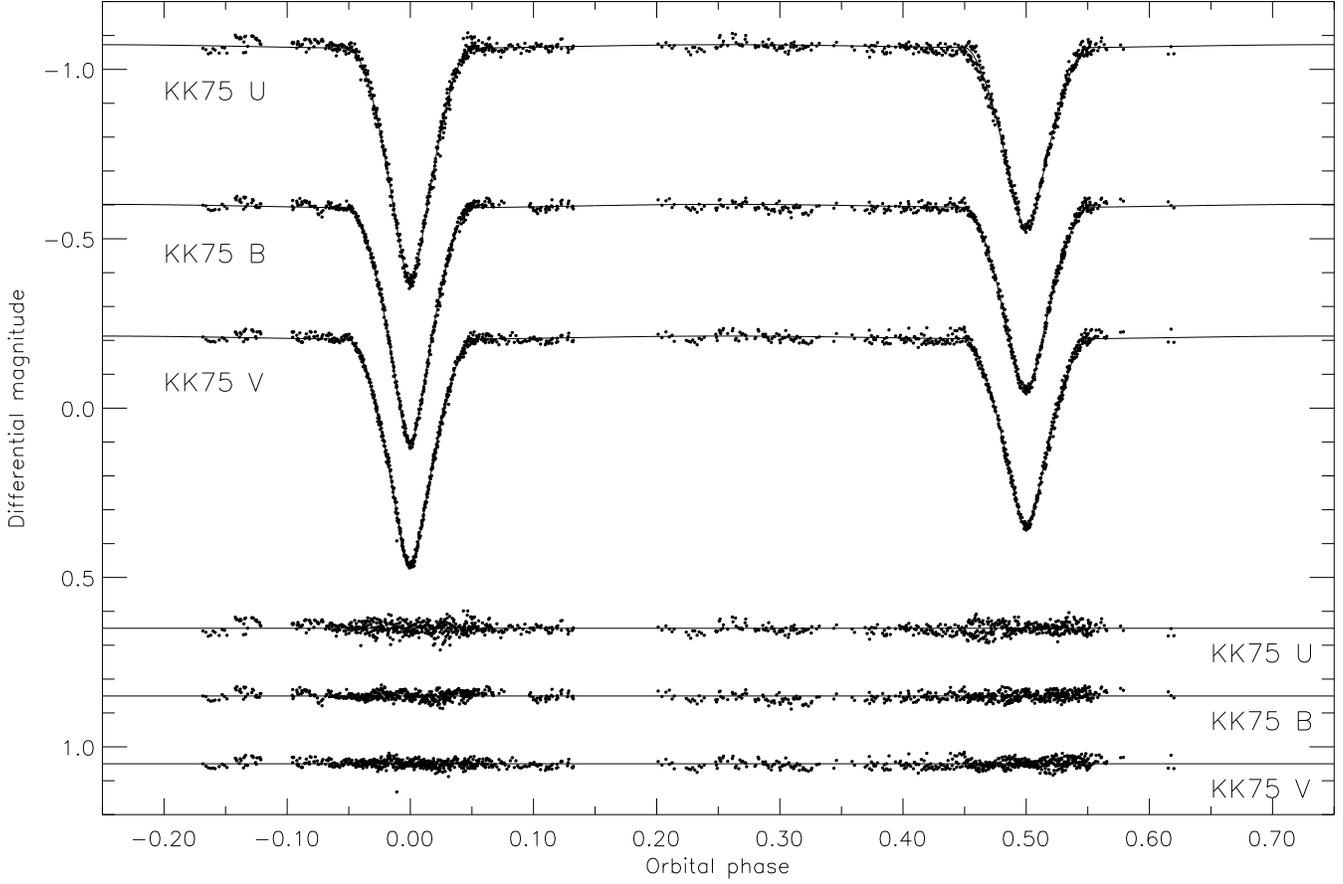} \\ 
\caption{\label{fig:lcplot1} The KK75 differential light curves of WW\,Aur, compared to the best-fitting light curves found using {\sc ebop}. The residuals of the fits are plotted with magnitude offsets for clarity.} \end{figure*}

Seven good photoelectric light curves of WW\,Aur exist: the $uvby$ observations obtained by E75 and the $UBV$ data of KK75. These were analysed with {\sc ebop}\footnote{Eclipsing Binary Orbit Program (written by PBE) \\ {\tt http://mintaka.sdsu.edu/faculty/etzel/}.} (E75; Popper \& Etzel 1981; Etzel 1981), which uses the Nelson-Davis-Etzel model (Nelson \& Davis 1972; Etzel 1981) in which the discs of stars are approximated using biaxial ellipsoids. {\sc ebop} is applicable only to eclipsing binaries which are sufficiently well detached for the stellar shapes to be close to spherical, as in the case of WW\,Aur. The version of {\sc ebop} used here has been modified to fit for the sum and ratio of the stellar radii and to use the Levenberg-Marquardt minimisation algorithm ({\sc mrqmin} from Press et al.\ 1992) to find the least-squares best fit to a light curve. 

Limb darkening was incorporated using the linear law (e.g., Van Hamme 1993) as this meant that the coefficients, $u_{\rm A}$ and $u_{\rm B}$, could be found from the observations. Using a more sophisticated non-linear law would have required us to adopt theoretically-calculated coefficients, so our photometric results would no longer have been entirely empirical. The gravity darkening exponents, $\beta_1$\footnote{The meaning of the  symbol $\beta_1$ is here given by $F \propto {\Teff}^4 \propto g^{\beta_1}$ where $F$ and $g$ are the local bolometric flux and gravity.}, were fixed at 1.0 (Claret 1998) and the mass ratio was fixed at the spectroscopic value; changes in both quantities have a negligible effect on the solutions. Third light and orbital eccentricity were found to be negligibly different from zero so were fixed at this value for the final light curve solutions.

\begin{figure*} \includegraphics[width=\textwidth,angle=0]{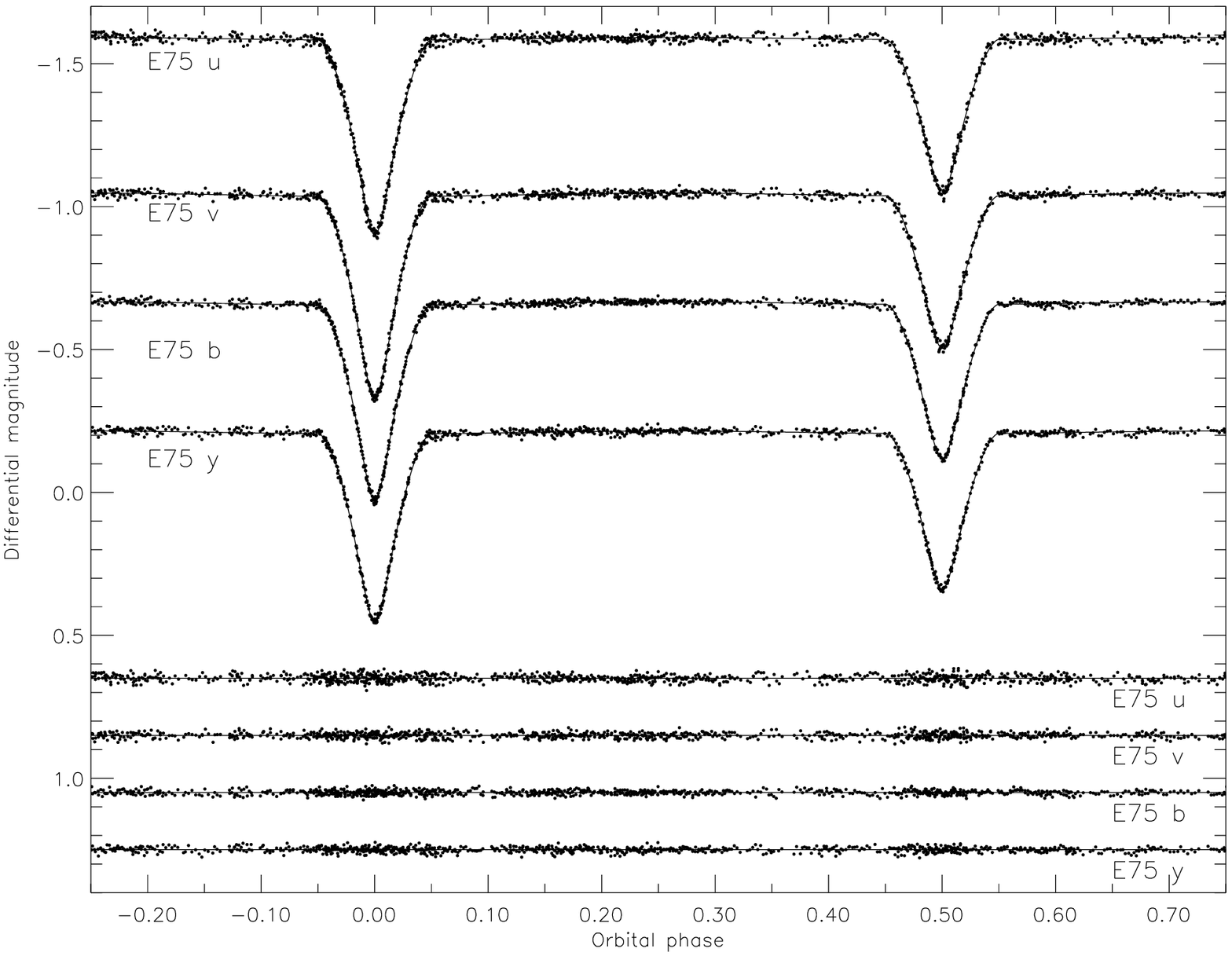} \\ 
\caption{\label{fig:lcplot2} The E75 differential light curves of WW\,Aur, compared to the best-fitting light curves found using {\sc ebop}. The residuals of the fits are plotted with magnitude offsets for clarity} \end{figure*}

The seven light curves were individually fitted using {\sc ebop} and allowing the stellar radii, surface brightnes ratio, orbital inclination and limb darkening coefficients to be adjusted. The results are given in Table~\ref{table:lcplot} and the best fits are plotted in Fig.~\ref{fig:lcplot1} and Fig.~\ref{fig:lcplot2}.

A solution was also obtained for the light curve obtained by Huffer \& Kopal (1951) using an unfiltered photoelectric photometer equipped with two different amplifiers. This solution is in very poor agreement with the other solutions and the data are not homogeneous, so was not considered further.

\begin{table*} \begin{center} 
\caption{\label{table:lcplot} Results of the light curve analysis for WW\,Aurigae. The uncertainties in individual parameters were found using Monte Carlo simulations. The adopted value and uncertainty for each parameter is the weighted mean and the standard error of the mean.} 
\begin{tabular}{l rrrrr@{\ \ \ \ \ \ \ \ }rrr@{\ \ \ \ \ \ \ \ }r} \hline
                                          &     &    \multicolumn{4}{c}{E75}        & \multicolumn{3}{c}{KK75} & Adopted\\
                                          &     &  $u$   & $v$    & $b$    & $y$    & $U$    & $B$    & $V$    & values\\ \hline
Number of datapoints                      &     & 981    & 962    & 902    & 903    & 999    & 980    & 1058   & 6785   \\
Observational scatter (mag)               &     & 0.012  & 0.010  & 0.008  & 0.009  & 0.016  & 0.012  & 0.012  &        \\[1pt]
Fractional total radii of the stars       &     & 0.3084 & 0.3101 & 0.3100 & 0.3099 & 0.3084 & 0.3092 & 0.3118 & 0.3099 \\
($r_{\rm A} + r_{\rm B}$)                 &$\pm$& 0.0012 & 0.0008 & 0.0009 & 0.0009 & 0.0011 & 0.0009 & 0.0009 & 0.0004 \\[1pt]
Ratio of the radii                        &     & 0.978  & 0.967  & 0.915  & 0.968  & 0.982  & 0.950  & 0.996  & 0.953  \\
($k$)                                     &$\pm$& 0.057  & 0.023  & 0.020  & 0.034  & 0.032  & 0.027  & 0.041  & 0.011  \\[1pt]
Fractional radius of primary star         &     & 0.1559 & 0.1577 & 0.1619 & 0.1575 & 0.1556 & 0.1585 & 0.1562 & 0.1586 \\
($r_{\rm A}$)                             &$\pm$& 0.0046 & 0.0019 & 0.0017 & 0.0028 & 0.0027 & 0.0023 & 0.0034 & 0.0009 \\[1pt]
Fractional radius of secondary star       &     & 0.1525 & 0.1524 & 0.1481 & 0.1524 & 0.1528 & 0.1506 & 0.1556 & 0.1515 \\
($r_{\rm B}$)                             &$\pm$& 0.0041 & 0.0018 & 0.0018 & 0.0026 & 0.0023 & 0.0022 & 0.0029 & 0.0009 \\[1pt]
Orbital inclination (\degr)               &     & 87.66  & 87.51  & 87.59  & 87.37  & 87.46  & 87.71  & 87.64  & 87.55  \\
($i$)                                     &$\pm$& 0.20   & 0.10   & 0.14   & 0.11   & 0.11   & 0.11   & 0.11   & 0.04   \\[1pt]
Surface brightness ratio                  &     & 0.862  & 0.828  & 0.861  & 0.884  & 0.761  & 0.817  & 0.869  &        \\
($J$)                                     &$\pm$& 0.041  & 0.030  & 0.029  & 0.032  & 0.026  & 0.027  & 0.027  &        \\[1pt]
Primary limb darkening coefficient        &     & 0.369  & 0.675  & 0.643  & 0.533  & 0.709  & 0.616  & 0.416  &        \\
($u_{\rm A}$)                             &$\pm$& 0.088  & 0.057  & 0.058  & 0.065  & 0.059  & 0.056  & 0.060  &        \\[1pt]
Secondary limb darkening coefficient      &     & 0.457  & 0.722  & 0.658  & 0.575  & 0.452  & 0.512  & 0.418  &        \\
($u_{\rm B}$)                             &$\pm$& 0.126  & 0.082  & 0.074  & 0.086  & 0.096  & 0.078  & 0.083  &        \\[2pt]
Light ratio (assuming $k = 0.953 \pm 0.011$) &  & 0.757  & 0.737  & 0.776  & 0.790  & 0.768  & 0.774  & 0.788  &        \\
($\frac{l_{\rm B}}{l_{\rm A}}$)           &$\pm$& 0.019  & 0.019  & 0.020  & 0.020  & 0.020  & 0.019  & 0.019  &        \\[1pt]
\hline \end{tabular} \end{center} \end{table*}

\subsection{Monte Carlo analysis}

\begin{figure} \includegraphics[width=0.48\textwidth,angle=0]{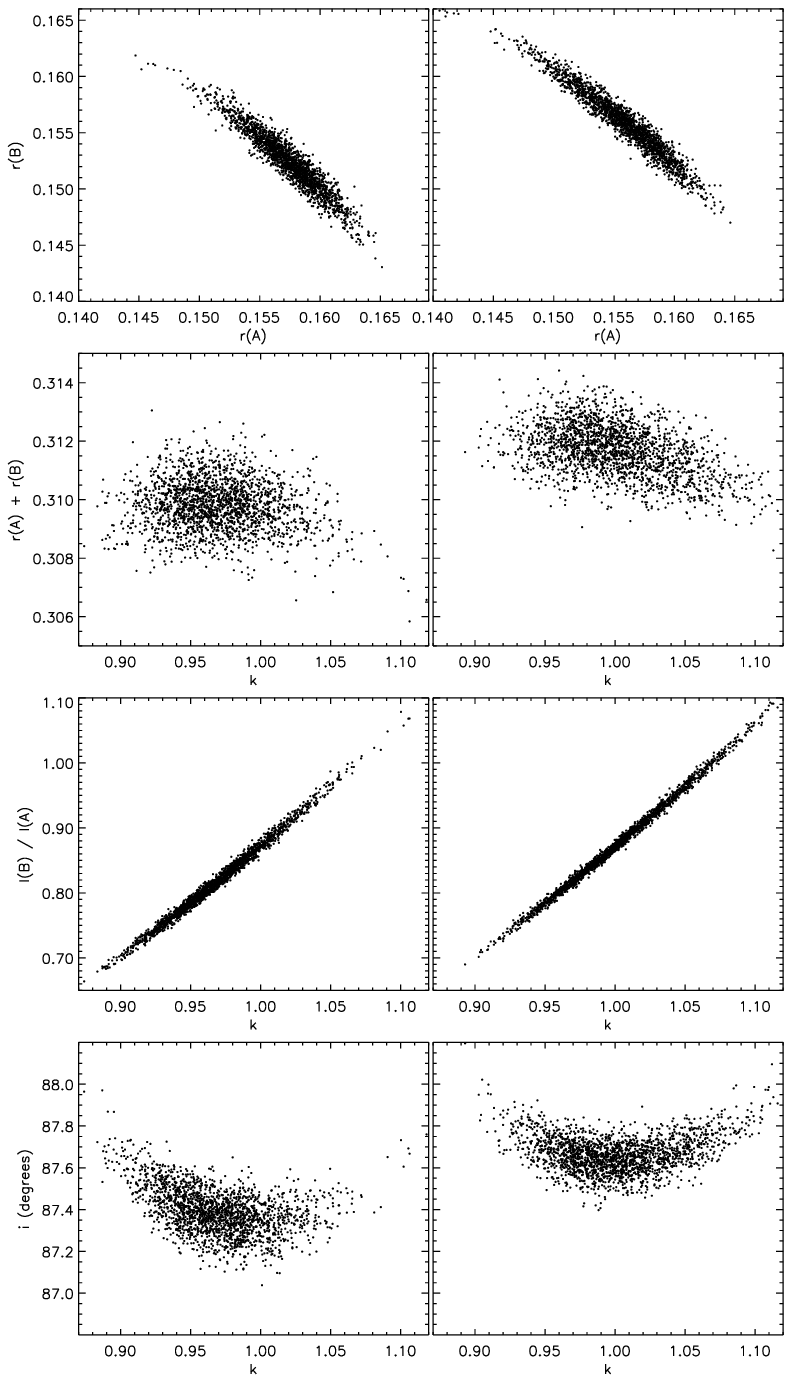} \\ 
\caption{\label{fig:lcerrs1} Sample distributions of the best-fitting parameters evaluated during the Monte Carlo analysis. The parameter symbols are as in Table~\ref{table:lcplot}. The $y$ (left-hand column) and $V$ (right-hand column) results are plotted as these passbands have a similar central wavelength but the light curves are from different sources.} \end{figure}

The formal uncertainties in parameters returned by minimisation algorithms in light curve modelling codes are known to be an inadequate representation of the overall parameter uncertainties (Popper 1984; Southworth et al.\ 2005), particularly when some parameters are strongly correlated. The ratio of the radii can be strongly correlated with the light ratio, and so the surface brightness ratio, in systems such as WW\,Aur, which have deep but not total eclipses. We have used Monte Carlo simulations to quantify the correlations between the photometric parameters (Southworth et al.\ 2004b, 2004c) and to derive robust uncertainties. For each observed light curve a `synthetic' light curve was constructed by evaluating the best-fitting model at the phases of observation. Then random Gaussian noise was added, of the same size as the scatter about the best fit, and a new best fit calculated. This process was undertaken ten thousand times for each observed light curve. 

The distributions of values of the best-fitting parameters from a Monte Carlo analysis allow the correlations between parameters to be investigated. For WW\,Aur, the ratio of the radii is strongly correlated with the surface brightness ratio but other correlations are minor. The most interesting results are shown in Fig.~\ref{fig:lcerrs1} for the $y$ and $V$ light curves. The standard deviation of the distribution of values for each parameter has been calculated and is given in Table~\ref{table:lcplot}. These standard deviations have been used to calculate the weighted mean and standard error of each parameter from the individual values from each light curve solution. The means and standard errors calculated in this way are in excellent agreement with values calculated just from the best-fitting parameters for each light curve. This confirms that the Monte Carlo analysis procedure provides robust and realistic uncertainties for parameters found by analysing light curves of dEBs (see also Southworth et al.\ 2004b).

\subsection{Limb darkening coefficients}

\begin{figure} \includegraphics[width=0.48\textwidth,angle=0]{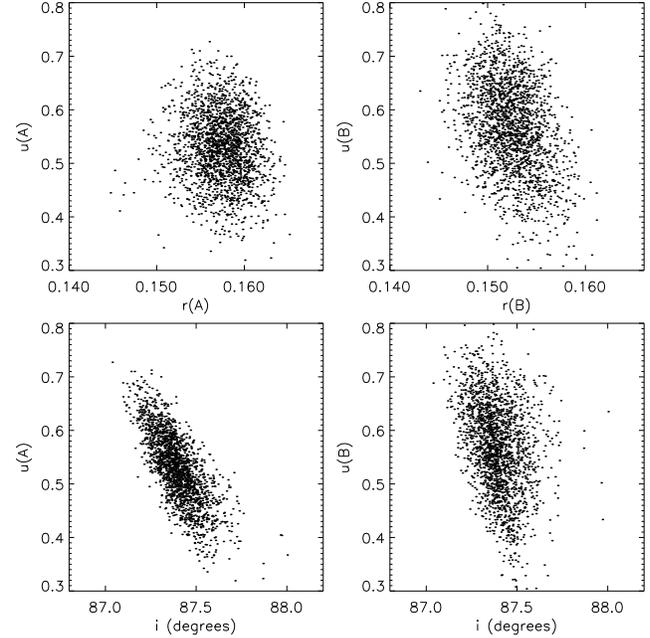} \\ 
\caption{\label{fig:lcerrs2} As Fig.~\ref{fig:lcerrs1} but concentrating on the limb darkening in the $y$ passband light curve (all panels).} \end{figure}

In our analysis of the light curves of WW\,Aur we fitted the linear limb darkening coefficients directly rather than fixing them at theoretically predicted values. We can therefore compare our best-fitting values with those calculated from theoretical model atmospheres, which are commonly used in the analysis of the light curves of dEBs. Fig.~\ref{fig:lcerrs2} shows that whilst $u_{\rm A}$ and $u_{\rm B}$ are not very well determined here, they are only weakly correlated with other photometric parameters.

\begin{figure} \includegraphics[width=0.48\textwidth,angle=0]{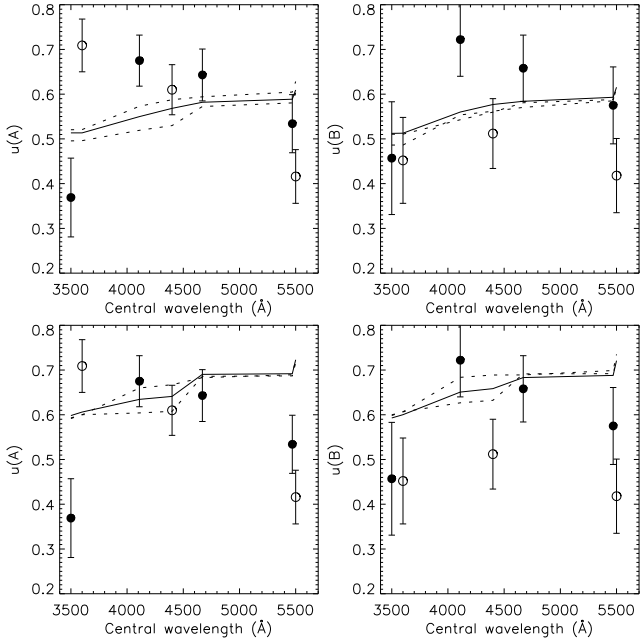} \\ 
\caption{\label{fig:LDcoeffs} The variation of the fitted limb darkening coefficients for the different passbands. The coefficients of the primary star are shown in the left-hand panels whilst those for the secondary are shown in the right-hand panels. Circles represent the coefficients found for WW\,Aur against the central wavelength of the passband used to observe that light curve. Filled circles represent the $uvby$ passbands and open circles the $UBV$ passbands. In the upper panels the theoretical coefficients of Van Hamme (1993) have been plotted using a dashed line (for the observed effective temperatures of the stars) and dotted lines (for the effective temperatures plus or minus their uncertainty). In the lower panels the coefficients of Claret (2000), for a metallicity of $\MoH = +0.5$, have been plotted.} \end{figure}

Fig.~\ref{fig:LDcoeffs} compares $u_{\rm A}$ and $u_{\rm B}$ to theoretical coefficients calculated by Van Hamme (1993) and Claret (2000). Coefficients for a high metallicity of $\MoH = +0.5$ have been chosen from Claret (2000); the coefficients of Van Hamme (1993) are available only for a solar chemical composition. Comparisons have been made at the central wavelengths of the $uvby$ (Str\"omgren 1963) and $UBV$ (Moro \& Munari 2000) passbands. Bilinear interpolation has been used to derive theoretical coefficients for the effective temperatures and surface gravities of the two stars from the tabulated coefficients; these are plotted for the temperatures of the stars and for the temperatures plus or minus their uncertainties.

The agreement between the observed limb darkening coefficients and the values derived using theoretical model atmospheres is generally reasonable, although the Claret (2000) coefficients are somewhat larger than those of Van Hamme (1993) and so have slightly worse agreement with the observations. It is important to remember, however, that the linear limb darkening law represents the limb darkening of stars significantly less well than other, more complex, limb darkening laws (Van Hamme 1993).

\subsection{Confidence in the photometric solution}

Values found for the ratios of the radii of stars in dEBs can depend on the model used to analyse the light curve (J.\ V.\ Clausen, 2004, private communication), due to different ways of representing the shapes of the stars and treating surface phenomena such as limb darkening. The $y$ light curve of WW\,Aur has also been analysed using the {\sc wink} code (Wood 1971). The resulting photometric solution is in good agreement with the solution determined using {\sc ebop}, suggesting that any systematic effects present in the {\sc ebop} solution are negligible. 

It is also useful to compare the photometric solution found here with the original solutions of the light curves. The solution of the $uvby$ light curves by Etzel (1975), who used the original version of the {\sc ebop} code, is in excellent agreement with the solution found here. As the version of {\sc ebop} used in this work is heavily modified, this provides confirmation that the modifications have not adversely affected our results. The original solution of the $UBV$ light curves (KK75) was obtained using rectification (see Section~\ref{sec:intro}) and is in good agreement with the results presented here.

The light ratios found in our light curve analysis agree well with a light ratio obtained from spectroscopic observations of the H$\beta$ line (see below). Whilst this is a useful consistency check, it is of limited importance here because the light ratio has a large error. More precise light ratios can be found from metallic lines, but these are unreliable as both components of WW\,Aur have peculiar spectra.

\subsection{Photometric indices}

The ratio of the radii determined from analysis of the seven light curves is $k = 0.953 \pm 0.011$. This ratio of the radii was used to determine the light ratios of WW\,Aur in the $uvby$ and $UBV$ passbands (Table~\ref{table:lcplot}). Str\"omgren photometric indices were then found for the two stars from the $uvby$ light ratios and the  indices of the overall system (Table~\ref{table:wwaurdata}). The reason for finding the light ratios using the same ratio of the radii for each light curve is that the resulting values are more directly comparable to each other as the statistical variation of the individual light curve solutions has been removed. 

The resulting Str\"omgren indices are given in Table~\ref{table:uvby}. The uncertainties in these indices have been calculated by adding in quadrature the effects of perturbing each input quantity by its own uncertainty. We have assumed that the $uvby$ filters used by E75 provide a good representation of the standard system.


\section{Effective temperature determination}                                                            \label{sec:teffs}  

\begin{table*} \begin{center} \caption{\label{table:uvby} $uvby$ photometry and atmospheric parameters for the combined system and for the individual stars.}
\begin{tabular}{lccccc} \hline
            &       $b-y$       &       $m_1$       &       $c_1$       &     \Teff      &     \logg        \\ \hline
System      & 0.081 $\pm$ 0.008 & 0.231 $\pm$ 0.011 & 0.944 $\pm$ 0.011 & 8210 $\pm$ 120 & 4.20 $\pm$ 0.05  \\
Primary     & 0.073 $\pm$ 0.019 & 0.215 $\pm$ 0.032 & 0.981 $\pm$ 0.031 & 8280 $\pm$ 300 & 4.13 $\pm$ 0.12  \\
Secondary   & 0.092 $\pm$ 0.023 & 0.252 $\pm$ 0.040 & 0.896 $\pm$ 0.041 & 8120 $\pm$ 340 & 4.29 $\pm$ 0.15  \\ \hline
\end{tabular} \end{center} \end{table*}

Fundamental values for the effective temperatures of the components of WW\,Aur have previously been found by Smalley et al.\ (2002), using the Hipparcos parallax of the system and ultraviolet, optical and infrared fluxes. We have repeated their analysis using our new values for the stellar radii and $V$-band magnitude difference. We obtain effective temperatures of $7960 \pm 420$\,K for star A and $7670 \pm 410$\,K for star B. These are only slightly different to previous results. The main contribution to the uncertainties from the Hipparcos parallax, with the lack of high-quality optical fluxes contributing to a slightly lesser extent. These fundamental values are in agreement with those obtained from H$\alpha$ and H$\beta$ profiles (Smalley et al.\ 2002). 

Ribas et al.\ (1998) determined the effective temperatures of the components of twenty dEBs from consideration of their Hipparcos parallaxes, apparent magnitudes and radii. The effective temperatures found by these authors for WW\,Aur are $8180 \pm 425$\,K and $7766 \pm 420$\,K, in good agreement with our values.

Str\"omgren $uvby$ photometry for the combined system and inferred values for the individual components (based on the light ratios determined in Section~\ref{sec:ebop}) are given in Table~\ref{table:uvby}, along with effective temperatures and surface gravities obtained from the solar-composition Canuto-Mazzitelli grids of $uvby$ colours (Smalley \& Kupka 1997). These imply slightly hotter temperatures and also a smaller difference between the two components. However, the fundamental values are consistent to within the uncertainties and thus will be preferred.

A light ratio was also obtained using the H$\beta$ line, which is not significantly affected by spectral peculiarities, to provide an external check on the accuracy of the light curve solution. Synthetic spectra were calculated using {\sc uclsyn} (Smith 1992; Smalley, Smith \& Dworetsky 2001), Kurucz (1993) {\sc atlas9} model atmospheres and absorption lines from the Kurucz \& Bell (1995) linelist. The spectra were rotationally broadened as necessary and instrumental broadening was applied with FWHM\,=\,0.2\,\AA\ to match the resolution of the observations. Comparison between synthetic spectra and observed spectra of WW\,Aur gave a light ratio of $\frac{l_2}{l_1} = 0.75 \pm 0.10$, which is in good agreement with the light ratios found by analysing the light curves. 

Projected rotational velocities of $35 \pm 3$ and $37 \pm 3$\kms, for stars A and B respectively, were obtained from the sodium D lines of WW\,Aur in archive MUSICOS spectra (Catala et al.\ 1993). These values are in excellent agreement with those found by Kitamura et al.\ (1976).


\section{Absolute dimensions}                                                                   \label{sec:dimensions}

The absolute dimensions of WW\,Aur, found from our spectroscopic and photometric analysis, are given in Table~\ref{table:dimensions}. We have adopted the usual convention of quoting standard errors when studying eclipsing binary stars. The results of our analyses are in good agreement with those of Kitamura et al.\ (1976).

WW\,Aur has a circular orbit and the rotational velocities of both components are synchronous to within the observational uncertainties, so a consideration of theories of tidal evolution is interesting. The timescales for orbital circularisation and rotational synchronisation due to tidal effects have been calculated using the theory of Zahn (1977, 1989) and the hydrodynamical mechanism of Tassoul (1987, 1988). The computations were performed using the same method as in Claret, Gim\'enez \& Cunha (1995) and Claret \& Cunha (1997). For the theory of Zahn, the critical times of orbital circularization for the system and rotational synchronisation for components A and B are 1380, 1120 and 1280\,Myr, respectively. For the Tassoul theory, the critical times are 250, 21 and 22\,Myr, respectively. The Tassoul timescales are much shorter than the Zahn timescales, as is usually found in similar studies. The physical basis of the Tassoul theory remains controversial (Claret \& Cunha 1997).

\begin{table} \begin{center} 
\caption{\label{table:dimensions} The absolute dimensions and related quantities determined for the detached eclipsing binary WW\,Aur. \Veq\ and \Vsync\ are the observed equatorial and calculated synchronous rotational velocities, respectively.}
\begin{tabular}{l r@{\,$\pm$\,}l r@{\,$\pm$\,}l} \hline
                                    & \multicolumn{2}{c}{WW\,Aur A} & \multicolumn{2}{c}{WW\,Aur B} \\ \hline
Mass (\Msun)                        & 1.964     & 0.007     & 1.814     & 0.007     \\
Radius (\Rsun)                      & 1.927     & 0.011     & 1.841     & 0.011     \\
\logg\ (\cms)                       & 4.162     & 0.007     & 4.167     & 0.007     \\ 
Effective temperature (K)           & 7960      & 420       & 7670      & 410       \\
$\log (L / L_\odot)$                & 1.129     & 0.092     & 1.023     & 0.093     \\ 
\Veq\ (\kms)                        & 35        & 3         & 37        & 3         \\
\Vsync\ (\kms)                      & 38.62     & 0.23      & 36.90     & 0.23      \\
\hline \end{tabular} \end{center} \end{table}


\section{Comparison with theoretical models}                                                    \label{sec:models}

\begin{figure*} \includegraphics[width=\textwidth,angle=0]{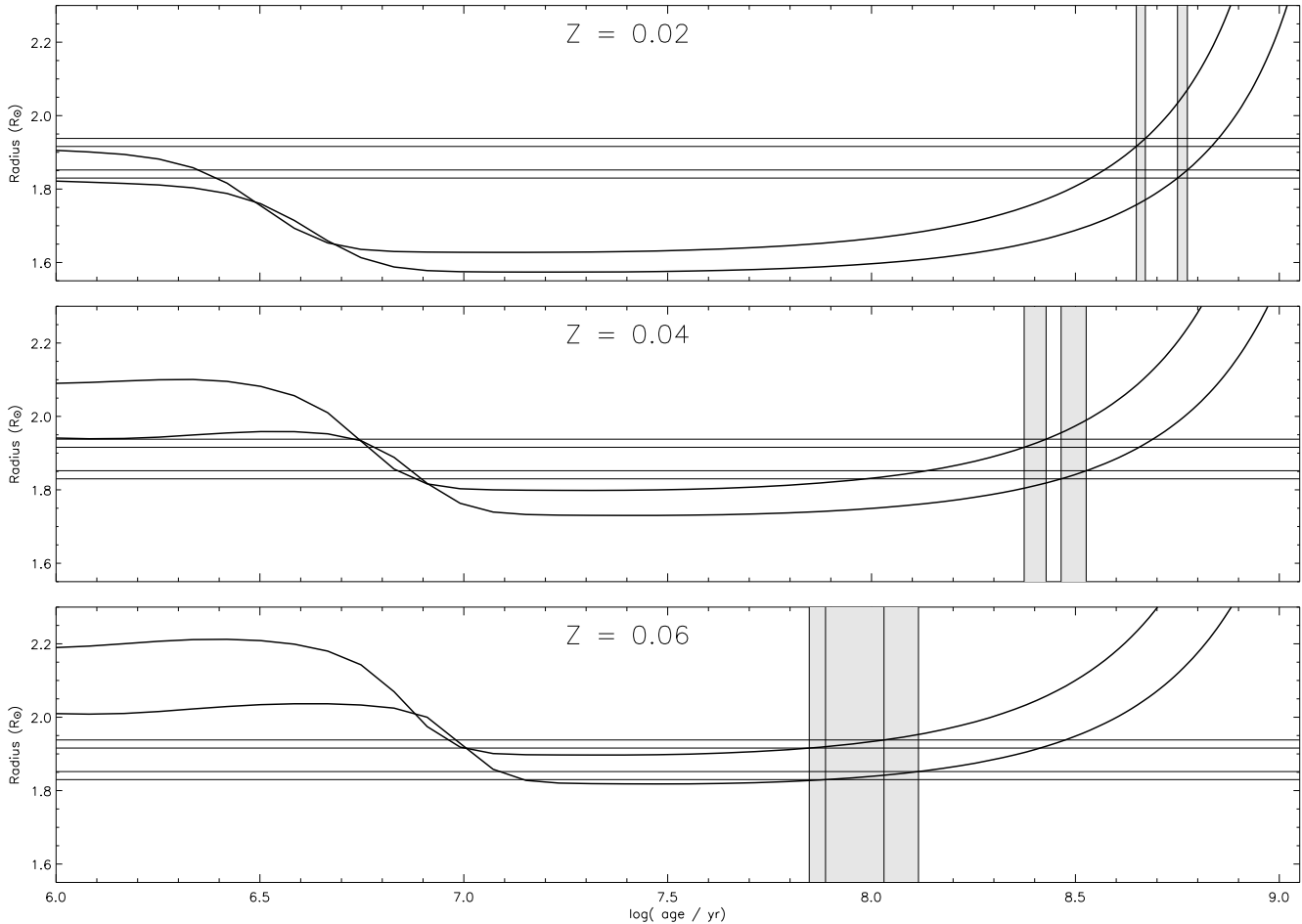} \\ 
\caption{\label{fig:modelfit} Comparison between the radii of the components of WW\,Aur and the predictions of the Claret (2004) theoretical stellar evolutionary models. Models have been calculated for the observed masses of the components of WW\,Aur and the evolution of the stellar radii are shown for metal abundances of $Z = 0.02$, 0.04 and 0.06. The lines of constant radius show the observed radii of WW\,Aur perturbed by their 1\,$\sigma$ uncertainties. Thick lines represent theoretical predictions, and the ages where these match the observed radii are shown by shaded areas.} \end{figure*}

We compared the masses and radii of the components of WW\,Aur to predictions of the Granada (Claret 1995), Geneva (Schaller et al.\ 1992), Padova (Girardi et al.\ 2000) and Cambridge (Pols et al.\ 1998) theoretical stellar evolutionary models, but were unable to find a match for any of the published chemical compositions under the assumption that the stars have the same age and chemical composition. We also compared the properties of WW\,Aur to the theoretical models of Siess et al.\ (2000) to investigate whether one or both of the stars may be in a pre-main-sequence phase, but we remained unable to find a match to the observations.

Further investigation using the models of Claret (2004) revealed that an acceptable match to the masses and radii of WW\,Aur could be obtained for a very high metal abundance of $Z = 0.06$ and an age between 77 and 107\,Myr. Models were calculated for the observed masses of the stars, and a helium abundance of $Y = 0.36$ was adopted from the standard chemical enrichment law with $\Delta Y / \Delta Z = 2$. The effective temperatures of the stars are in agreement with the model predictions for this chemical composition and age, although this is of low significance because they are quite uncertain. It is possible to match the properties of the stars using models with fractional metal abundances between about 0.055 and 0.065. An acceptable match to the observed properties of WW\,Aur can also be found for an age of about 1\,Myr and an approximately solar chemical composition, but the age range over which the match is acceptable is extremely small. These results are shown in Fig.~\ref{fig:modelfit}. 

We have investigated whether a lower chemical composition could be found by varying the input physics of the Claret (2004) models, which use nuclear reactions rates from NACRE. We have performed tests with the new measurement of the nuclear rate
$^{14}$N($p,\gamma$)$^{15}$O (Runkle 2003; Formicola et al.\ 2004) which is around a factor two smaller than the previous one (Schr\"oder et al.\ 1987). The differences are not noticeable for stars with masses similar to those of WW\,Aur. The influence of convective core overshooting was also investigated: for no extra mixing ($\alpha_{\rm OV} = 0$) we obtain slightly smaller ages but a very similar metal abundance. The effects of rotation were also considered (Claret 1999): for solid body rotation and local conservation of the angular momentum, no conspicuous differences were found when compared with standard models. 

We conclude that the high metal abundance required to match the physical properties of WW\,Aur cannot be attributed to differences among the evolutionary codes or the input physics. In fact, recent comparisons between the available grids of stellar models (Lastennet \& Valls-Gabaud 2002; Southworth et al.\ 2004b) indicate that the models from Geneva, Padova and Granada are very similar when compared with observations of dEBs.


\section{Discussion}                                                                            \label{sec:discussion}

The component stars of WW\,Aur are peculiar, both because of their metallic-lined nature and because current theoretical evolutionary models can only match their masses and radii for a high metal abundance of $Z \approx 0.06$. The strong metal lines in their spectra are due to the Am phenomenon, and not a high overall metal abundance, because the calcium and scandium lines are weak.

We must now ask whether the Am phenomenon is just a surface phenomenon, or whether it is affecting the radii of the two stars and causing us to find a spurious large metal abundance. This question can be answered by considering other well-studied metallic-lined dEBs and by investigating whether the diffusion of chemical species can significantly affect the radii of A stars predicted by theoretical models.

Several dEBs containing Am stars have recently been studied: V885\,Cyg (Lacy et al.\ 2004a), V459\,Cas (Lacy et al.\ 2004b), WW\,Cam (Lacy et al.\ 2002) and V364\,Lac (Torres et al.\ 1999). The components of V885\,Cyg and WW\,Cam are clearly metallic-lined but the presence of chemical peculiarity in V459\,Cas and V364\,Lac is uncertain. The theoretical models of Claret (2004) can match the properties of V885\,Cyg and WW\,Cam for metal abundances of 0.030 and 0.020, respectively. The same models can also match V459\,Cas with $Z = 0.012$ and V364\,Lac with $Z = 0.020$. As these dEBs can be matched by models with normal chemical compositions, we have no reason to suppose that the Am phenomenon makes a significant difference to the radii of stars (see Southworth et al.\ 2004d for further discussion). The metallic-lined dEB KW\,Hya was analysed by Andersen \& Vaz (1984, 1987), who found an unusual chemical composition from comparison with the models of Hejlesen (1980). However, the Hejlesen models use old opacity data which is quite different from recent results, so any chemical composition derived using them needs revision.

Theoretical models for A stars with diffusion have been published by Richer, Michaud \& Turcotte (2000). The Am phenomenon is found to be important only near the surfaces of A stars, so only affects surface quantities such as effective temperature (S.\ Turcotte, 2004, private communication). It will therefore have a negligible effect on the stellar radius. In particular, for WW\,Aur to match theoretical models of a normal chemical composition, the {\em ratio} of the radii of the two stars must become significantly smaller. Any physical phenomenon must therefore affect one component far more than the other component.

We have found no evidence that the Am phenomenon is causing us to find a high metal abundance for WW\,Aur, so we will now consider the existence of stars which are very rich in metals. The highest recent estimation of the metal abundance of a dEB is $Z = 0.042$ (EW\,Ori, Popper et al.\ 1986) by Ribas et al.\ (2000), from a comparison with (extrapolated) predictions from the Claret (1995) theoretical models.

A metal abundance of $Z = 0.06$, corresponding to $\FeH = +0.5$, is at least three times higher than the accepted solar abundance (but see Asplund, Grevasse \& Sauval 2004). However, the metal abundance of the old open cluster NGC\,6253 has been found to be between $Z = 0.04$ and 0.06 (Twarog, Anthony-Twarog \& De Lee 2003). These authors found metallicities between $\FeH = 0.4$ and 0.9 from the Str\"omgren $m_1$ and calcium $hk$ photometric indices of stars in NGC\,6253, although the best isochronal match to the morphology of its colour-magnitude diagram was obtained using the Padova stellar evolutionary models for $Z = 0.04$ and enhanced abundances of the $\alpha$-elements. Therefore, whilst high metal abundances of $Z = 0.06$ are unusual, there is no reason to assume that they do not exist.


\section{Summary and conclusion}

We have studied the metallic-lined A-type dEB WW\,Aurigae in order to determine its physical properties. The masses have been derived to accuracies of 0.4\% by cross-correlating the observed spectra against standard star spectra. The radii have been derived to accuracies of 0.6\% by analysing $UBV$ and $uvby$ light curves with the {\sc ebop} code. The masses and radii have been found without using any theoretical calculations so can be regarded as being entirely empirical. They are also among the most accurately known for any stars. The effective temperatures of the two stars have been derived from their Hipparcos parallax, using a method which is nearly fundamental.

Attempts to find a good match between the physical properties of WW\,Aur and predictions from several sets of theoretical stellar evolutionary models were unsuccessful for any of the chemical compositions for which models are available. Pre-main sequence evolutionary predictions were equally unsuccessful. However, the predictions of the Claret (2004) models match the observed masses, radii and effective temperatures for a metal abundance of $Z = 0.06$ and for an age of roughly 90\,Myr. This conclusion is robust against changes in the input physics of the models.

The tidal evolution theory of Zahn predicts timescales for orbital circularisation and rotational synchronisation which are much longer than the age of WW\,Aur, in disagreement with the observed circular orbit and the synchronous rotation of the stars. Similar cases can be found in Southworth et al (2004a) and Gonz\'alez \& Lapasset (2002). The Tassoul theory gives much shorter tidal evolution timescales and correctly predicts synchronous rotation, but predicts that the orbit should not yet be circularised. However, this could be easily explained by tidal evolution in the pre-main sequence phase (Zahn \& Bouchet 1989) or the formation of a system with a nearly circular orbit (Tohline 2002).

We have found no evidence to suggest that the Am phenomenon is causing us to derive a spurious metal abundance for WW\,Aur. In fact, the Claret (2004) theoretical evolutionary models can match the observed properties of other Am dEBs for entirely normal chemical compositions. This is in agreement with the results of stellar models which include diffusion, which suggest that its presence does not significantly affect the radii of A stars. 

Our conclusion that WW\,Aur is rich in metals relies on our measurements of the masses and radii of the two stars and the use of theoretical models to determine what initial chemical compositions could produce these properties, so is reliable. Further investigation of this will require more accurate effective temperatures, which may be available in the near future. A final analysis of the astrometric data from the Hipparcos satellite will soon be available, which is expected to report a significantly more accurate parallax for this dEB (F.\ van Leeuwen, 2004, private communication; van Leeuwen \& Fantino 2005). WW\,Aur will also be a target for the ASTRA robotic spectrophotometer (Adelman et al.\ 2004), which aims to observe accurate spectrophotometric fluxes for the measurement of fundamental effective temperatures of many types of stars. A rediscussion of the effective temperatures of WW\,Aur compared to theoretical predictions may then allow chemical composition of the system to be studied in more detail.


\section*{Acknowledgements}

Many thanks are due to Bill Morris for technical help with retrieving electronic versions of the $uvby$ light curves from a nine-track tape. Thanks are also due to Jens Viggo Clausen for analysing the $y$ light curve of WW\,Aur with {\sc wink}. We would also like to thank Jens Viggo Clausen, Sylvain Turcotte, Johannes Andersen, J{\o}rgen Christensen-Dalsgaard and many others for discussions.

The following internet-based resources were used in research for this paper: the ESO Digitized Sky Survey; the NASA Astrophysics Data System; the SIMBAD database operated at CDS, Strasbourg, France; the VizieR service operated at CDS, Strasbourg, France; and the ar$\chi$iv scientific paper preprint service operated by Cornell University. 

JS acknowledges financial support from PPARC in the form of a postgraduate studentship. The authors acknowledge the data analysis facilities provided by the Starlink Project which is run by CCLRC on behalf of PPARC.  

This paper is based on observations made with the Isaac Newton Telescope operated on the island of La Palma by the Isaac Newton Group in the Spanish Observatorio del Roque de los Muchachos of the Instituto de Astrof\'\i sica de Canarias.

This publication makes use of data products from the Two Micron All Sky Survey, which is a joint project of the University of Massachusetts and the Infrared Processing and Analysis Center/California Institute of Technology, funded by the National Aeronautics and Space Administration and the National Science Foundation.


\end{document}